\documentclass{article}
\usepackage{ws-crop961x669}
\usepackage{calligra}
\usepackage{amsthm}
\usepackage{amsmath}
\usepackage{amssymb}
\usepackage{amscd}
\usepackage{bbm}
\usepackage[all]{xy}
\usepackage{mathrsfs}

\def\L{\mathscr{L}}
\def\id{\mathrm{id}}
\def\YMSM{\mathrm{YMSM}}
\def\YM{\mathrm{YM}}

\def\F{\mathfrak{F}}
\def\f{\mathfrak{f}}
\newcommand{\R}{\mathbb{R}}

\newcommand{\C}{\mathbb{C}}

\newcommand{\Ad}{\mathrm{Ad}}
\newcommand{\ad}{\mathrm{ad}}
\newcommand{\Hor}{\mathrm{Hor}}
\newcommand{\Ker}{\mathrm{Ker}}
\newcommand{\inv}{\mathrm{inv}}
\newcommand{\altt}{\mathrm{alt}}
\newcommand{\triv}{\mathrm{triv}}

\newcommand{\Mor}{\mathrm{Mor}}

\title{Yang--Mills--Scalar--Matter Fields in the Two-Point Space }
\author{ Gustavo Amilcar Salda\~na Moncada \\ Instituto de Matem\'aticas, UNAM,\\ Area de la Investigaci\'on Cientifica, Circuito Exterior,\\ Ciudad Universitaria, Ciudad de M\'exico, CP 04510, M\'exico.  \\ e-mail: gamilcar@matem.unam.mx  }
\begin{document}
\maketitle
\begin{abstract}
The purpose of this work is to present a Non--Commutative Geometrical version of the Yang--Mills Theory and the Yang--Mills Scalar Matter Theory by constructing a concrete example using M. Durdevich's theory of quantum principal bundles.
\end{abstract}
\section{Introduction}
The study of principal bundles and principal connections have applications in Topology and Differential Geometry among other areas. In Physics, it provides the natural framework to talk about Classical and Quantum Field Theory; in particular, it is possible to talk about the Yang--Mills theory and the theory of matter fields. For the purpose of this work, we will consider that for a Yang--Mills Scalar Matter model it is necessary:

\begin{enumerate}
\item A Riemannian or a Pseudo--Riemannian closed and oriented manifold $(M,g)$.
\item A principal $G$--bundle over $M$, $\pi:GM\longrightarrow M$ with $G$ a compact group. 
\item An $\ad$--invariant inner product in the Lie algebra $\mathfrak{g}$ of $G$.
\item A unitary representation $\alpha$ of $G$ acting on $V^\alpha$
\item A smooth function $V:\R\longrightarrow \R$ called {\it the potential}. 
\end{enumerate}
Using these objects we define the {\it Lagrangian} 
\begin{equation}
\label{f.1}
\L_{\YMSM}(\omega,T):=-\dfrac{1}{2}\langle R^\omega,R^\omega\rangle+\dfrac{1}{2}\left(\langle \nabla^{\omega}_\alpha T,\nabla^{\omega}_\alpha T\rangle-V(T)\right)
\end{equation}

\noindent where $\omega$ is a principal connection, $R^\omega$ is its curvature viewed as $\mathfrak{g}M$--valued $2$--form, $T$ $\in$ $\Gamma(M,V^\alpha M)$, $\nabla^{\omega}_\alpha$ is the induced linear connection and $V(T)=V(\langle T,T\rangle)$. 

This Lagrangian is gauge--invariant and the critical points of its associated action 
\begin{equation}
\label{f.2}
\mathscr{S}_{\YMSM}(\omega,T)=\int_M \mathscr{L}_{\YMSM}(\omega,T)\;dvol_g
\end{equation} 
satisfy

\begin{equation}
\label{f.2.1}
\langle\nabla^\omega_\alpha T\mid\lambda\rangle  =\langle d^{\nabla^\omega_{\ad}\,\star} R^\omega\mid\lambda \cdot T\rangle 
\end{equation}
for all $\mathfrak{g}M$--valued $1$--form  $\lambda$, where $d^{\nabla^\omega_{\ad}\,\star}$ is the adjoint operator of the exterior covariant derivative of the induced linear connection $\nabla^\omega_{\ad}$; and
\begin{equation}
\label{f.2.2}
{\nabla^\omega_{\alpha}}^\star \nabla^\omega_{\alpha} T-V'(T)=0.
\end{equation}
These equations describe the dynamics of {\it scalar matter particles coupled to gauge bosons} with symmetry $G$.

In this way, it should be natural to thing that a non--commutative geometrical version of the Yang--Mills theory and scalar matter fields has to arise from the non--commutative geometrical version of the theory of principal bundles and principal connections.

M. Durdevich developed in \cite{micho1}, \cite{micho2} (among other works) a complete theory of quantum principal bundles (qpbs) and quantum principal connections (qpcs) using the concept of {\it quantum group} studied by S. Woronovicz in \cite{woro1}, \cite{woro2} as the structure group of the bundle.  Durdevich's theory uses the universal differential envelope $\ast$--calculus as {\it quantum differential forms} on the group; however, every differential $\ast$--calculus that extends the comultiplication can be used, for example, the braided exterior $\ast$--calculus. The universal differential envelope $\ast$--calculus is maximal with this property. If the read wants to know in detail the theory, we recommend the book written by S. Sontz \cite{stheve}, where the reader can find explicit calculations and exercises.

Using Durdevich's theory one can get a very natural {\it non--commutative geometrical} version of the Yang--Mills theory and the Yang--Mills Scalar Matter theory, i.e., one can get the {\it non--commutative geometrical} version of Equations (2), (3) and (4) \cite{sald1}, \cite{sald2}, \cite{sald3}. This work aims is to show a concrete example of this theory. Specifically, we will work on the quantum trivial principal bundle given by the two--point space as the base space and the symmetric group of order $2$, $S_2$, as the structure group; and on this quantum principal bundle we will show the respective Lagrangians and the {\it non--commutative geometrical} associated field equations, as well as their solutions. It is worth mentioning again that this is just one example of the general theory.

To accomplish  our purpose this paper breaks down into $9$ parts. In the second one we present the two--point space in the Non--Commutative Geometry's framework, the graded differential $\ast$--algebra that we will use as well as some other geometrical concepts, for example, the codifferential. In the third one we will talk about the structure quantum group $S_2$. In the fourth section we are going to present the quantum principal bundle on which we will work as well a characterization of its quantum principal connections, curvatures and covariant derivatives. In the fifth section we will present the theory of associated quantum vector bundles which includes induced quantum linear connections, exterior covariant derivatives and their adjoint operators. The sixth section is to present our definition of the quantum gauge group which is based on the one presented by T. Brzezinski in \cite{Br}. The seventh section is for the {\it non--commutative geometrical} version of the Yang--Mills theory and the eighth one is for the {\it non--commutative geometrical} version of the Yang--Mills Scalar Matter theory. In the last section we will talk about some final and important comments.

Unfortunately, I did not meet Zbigniew Oziewicz, but I have heard to many stories about him, particularly, because he was a regular loved member of the Seminar of Quantum Geometry directed by M. Durdevich at UNAM. He used to work in Category Theory and I think that the categorical formulation of this theory presented in \cite{sald2} would have liked him.  

\section{The Two--Point Space}

Let us start taking the two--point space $\{x_0,x_1 \}$ and the {\it quantum space} ($C^\ast$--algebra)
\begin{equation*}
(M:=C_\C(\{x_0,x_1 \}),\cdot,\mathbbm{1}(x),||\,\,||_\infty,\ast).
\end{equation*}
A canonical basis of $M$ as $\C$--vector space is given by the functions 
\begin{equation*}
\beta_M:=\{\{x_0,x_1 \}:M\longrightarrow \C\,,\quad  p_1:\{x_0,x_1 \}\longrightarrow \C\},
\end{equation*}
which are defined by
\begin{equation*}
p_0(x)=\left\{ 0 \;\;\mbox{ if }\;\; x=x_1
\atop
1 \;\; \mbox{ if }\;\; x=x_0,
\right. \qquad p_1(x)=\left\{ 0 \;\;\mbox{ if }\;\; x=x_0
\atop
1 \;\; \mbox{ if }\;\; x=x_1.
\right.
\end{equation*}

Our next step is considering the universal graded differential $\ast$--algebra of $M$ without $n$--forms for $n \geq 3$ 
\begin{equation*}
(\Omega^\bullet(M),d,\ast).
\end{equation*} 
We can represent this algebra as follows:
$$M \longleftrightarrow \left\lbrace p=\left(\begin{array}{lcr}
 z_0              &  0 \\
 0 & z_1 \\
\end{array}\right) \mid z_0,\,z_1 \,\in\, \C \right\rbrace,$$ $$\Omega^1(M) \longleftrightarrow \left\lbrace \mu=\left(\begin{array}{lcr}
 0              &  z_0 \\
 z_1 & 0 \\
\end{array}\right) \mid z_0,\,z_1 \,\in\, \C \right\rbrace,$$
$$\Omega^2(M) \longleftrightarrow \left\lbrace \mu=\left(\begin{array}{lcr}
 z_0              &  0 \\
 0 & z_1 \\
\end{array}\right) \mid z_0,\,z_1 \,\in\, \C \right\rbrace.$$ This space will play the role of {\it quantum} differential forms on $M$.

Consider 
\begin{equation*}
dvol:=p_0\,dp_1\, dp_0+ p_1\,dp_1\, dp_1 =\left(\begin{array}{lcr}
-i  &  0 \\
\hspace{0.1cm}0 & i \end{array}\right) \;\in\;\Omega^2(M). 
\end{equation*}
We can define left and right hermitian structures on $\Omega^\bullet(M)$: 

\begin{equation*}
\begin{aligned}
\langle-,-\rangle: M\times M&\longrightarrow M\\
(\;\hat{p}\;,\;p\;)&\longmapsto \hat{p}\,p^\ast;
\end{aligned}
\end{equation*}
\begin{equation*}
\begin{aligned}
\langle-,-\rangle: \Omega^1(M)\times \Omega^1(M)&\longrightarrow M\\
(\;\hat{\mu}\;,\;\mu\;)&\longmapsto \hat{\lambda}_0\,\lambda^\ast_0\,p_0+\hat{\lambda}_1\,\lambda^\ast_1\,p_1
\end{aligned}
\end{equation*}
if $\hat{\mu}=\left(\begin{array}{lcr}
0  &  \hat{\lambda}_0 \\
\hat{\lambda}_1 & 0 \end{array}\right),$ $\mu=\left(\begin{array}{lcr}
0  &  \lambda_0 \\
\lambda_1 & 0 \end{array}\right)$;
\begin{equation*}
\begin{aligned}
\langle-,-\rangle: \Omega^2(M)\times \Omega^2(M)&\longrightarrow M\\
(\;\hat{p}\,dvol\;,\;p\,dvol\;)&\longmapsto \hat{p}\,p^\ast.
\end{aligned}
\end{equation*}
and finally $$\langle\hat{\mu},\mu\rangle_r=\langle\hat{\mu}^\ast,\mu^\ast\rangle_l,$$  which will play the role of {\it left and right quantum riemannian structures} on $M$.
A {\it quantum} integral can be defined as the linear map
\begin{equation*}
\int_M : \Omega^2(M) \longrightarrow \C\; \mbox{ s.t. }\displaystyle \int_M p_0\,dp_1\,dp_0=\dfrac{1}{2}=\int_M p_1\,dp_1\,dp_1.
\end{equation*}
All of these allow us to introduce the inner products $$\langle-|-\rangle_l=\int_M \langle-,-\rangle_l\,dvol\,,\quad  \langle-|-\rangle_r=\int_M \langle-,-\rangle_r \,dvol.$$

The {\it left and right quantum Hodge operators} $\star_l$, $\star_r$ can be defined and with that, the {\it left and right codifferential}: $$d^{\star_l}\mu=\left(\begin{array}{lcr}
i(z_0-z_1)  & \hspace{0.2cm} 0 \\
\hspace{0.8cm} 0 & -i(z_0-z_1) \end{array}\right)$$ for $\mu$ $\in$ $\Omega^1(M)$, $$d^{\star_l}\mu=\left(\begin{array}{lcr}
\hspace{0.8cm}0  &  -(z_0+z_1) \\
-(z_0+z_1) & 0 \end{array}\right) $$ for $\mu=$ $\in$ $\Omega^2(M)$ and $$d^{\star_r}:=\ast \circ d^{\star_l} \circ \ast.$$ Of course $$\langle \hat{\eta}|d\eta\rangle_l=\langle d^{\star_l}\hat{\eta}|\eta\rangle_l \quad \mbox{ and }\quad \langle \hat{\eta}|d\eta\rangle_r=\langle d^{\star_r}\hat{\eta}|\eta\rangle_r.$$ 

\section{The Group $S_2$}

The {\it quantum} space $(G:=C_\C(S_2),\cdot,\mathbbm{1}(x),||\,||_\infty,\ast)$ associated to the group $S_2$ is basically the same as $M$ but we will use the notation $$\{\Delta_0,\Delta_1 \}$$ for its basis. The structure of Hopf algebra will be denoted by $(\phi,\epsilon,\kappa)$, where $\phi$ is the coproduct, $\epsilon$ is the counity and $\kappa$ is the coinverse. Also we are going to use the universal graded differential $\ast$--algebra\footnote{It is worth mentioning that in this case, the universal differential envelope $\ast$--calculus agrees with the universal graded differential $\ast$--algebra.} $$(\Gamma^\wedge,d,\ast)$$ as {\it quantum} differential forms on $G$.\\

In this way the space ${_\inv}\Gamma:=\Ker(\epsilon)$ will play the role of the {\it quantum} Lie algebra and $$\{\varsigma:=(\Delta_0-\Delta_1) d\Delta_1 \}$$ is a basis of ${_\inv}\Gamma$.

\section{The Quantum Principal Bundle}

We will consider the trivial quantum principal bundle (qpb) $$\zeta:=(M\otimes G,M,\Phi),$$ where $\Phi(x\otimes g)=x\otimes \phi(g)$ \cite{stheve}.

 The differential calculus $(\Omega^\bullet(M\otimes G), d,\ast)$ on the bundle  will be the tensor product of $(\Omega^\bullet(M), d,\ast)$ and  $(\Gamma^\wedge,d,\ast)$. The extension of $\Phi$ will be given by $\Psi(\mu\otimes \vartheta)=x\otimes \phi(\vartheta),$ where here, $\phi$ is extension of the coproduct on $(\Gamma^\wedge,d,\ast)$ \cite{stheve}.
 
In this case, quantum principal connections (qpcs) $$\omega:{_\inv}\Gamma\longrightarrow \Omega^1(M\otimes G)$$ are completely defined by the equation $$\omega(\varsigma)=\mu\otimes \mathbbm{1}+\mathbbm{1}\otimes \varsigma, $$ for some $\mu$ $\in$ $\Omega^1(M)$ \cite{micho2}.

The curvature of any qpc $$R^{\omega}:{_\inv}\Gamma\longrightarrow \Omega^2(M\otimes G)$$ is defined by $$R^{\omega}(\varsigma)=d\mu-2\,\mu.\mu\;\in\;\Omega^2(M);$$ while the covariant derivative $$ D^\omega:\Hor^\bullet(M\otimes G)=\Omega^\bullet(M)\otimes G\longrightarrow\Hor^\bullet(M\otimes G)=\Omega^\bullet(M)\otimes G$$
is given by $$D^\omega(\eta\otimes g)=d(\eta\otimes g)-(-1)^{\partial \eta}\eta.\omega(\kappa(g^{(1)})dg^{(2)})$$ if $\phi(g)=g^{(1)}\otimes g^{(2)}$ (in Sweedler's notation) \cite{stheve}.

\section{Associated Quantum Vector Bundles}

For any unitary finite--dimensional corepresentation $\alpha$ on $V^\alpha$ of $G$, the space $\Mor(\alpha,\Phi)$ (morphisms of corepresentations) is a $M$--bimodule. Even more, it is a  finitely generated projective left and right $M$--module. Taking the left (right) structure, we define the {\it left associated quantum vector bundle} ({\it right associated quantum vector bundle}) as $\Gamma^l(M,V^\alpha M):=\Mor(\alpha,\Phi)$ ($\Gamma^r(M,V^\alpha M):=\Mor(\alpha,\Phi)$).

It is possible to define non--singular {\it hermitian structures} on these quantum vector bundles (qvbs) by means of
$$\langle T_1,T_2\rangle_l=\sum^n_{k=1}T_1(e_k)T_2(e_k)^\ast\,,\quad \langle T_1,T_2\rangle_r=\sum^n_{k=1}T_1(e_k)^\ast T_2(e_k),$$ where $\{e_k \}^n_{k=1}$ is an orthogonal basis of $V^\alpha$.

There exists a canonical isomorphism between $\Mor(\alpha,\Psi|_{\Hor^\bullet(M\otimes G)})$ and $\Omega^\bullet(M)\otimes_M \Mor(\alpha,\Phi)$ ({\it left qvb--valued  differential forms}). Moreover, there exists a canonical isomorphism between the space $\Mor(\alpha,\Psi|_{\Hor^\bullet(M\otimes G)})$ and $ \Mor(\alpha,\Phi)$ $\otimes_M \Omega^\bullet(M)$ ({\it right qvb--valued  differential forms}). Taking a qpc $\omega$ and denoting the last isomorphisms by $\Upsilon_{\alpha}$ and $\widehat{\Upsilon}_{\alpha}$ respectively, the linear maps $$\nabla^{\omega}_{\alpha}:=\Upsilon_{\alpha}\circ D^{\omega}:\Gamma^l(M,V^\alpha M)\longrightarrow \Omega^1(M)\otimes_M \Gamma^l(M,V^\alpha M) $$ $$\widehat{\nabla}^{\omega}_{\alpha}:=\widehat{\Upsilon}_{\alpha}\circ \ast \circ D^{\omega}\circ \ast: \Gamma^r(M,V^\alpha M) \longrightarrow \Gamma^r(M,V^\alpha M)\otimes_M \Omega^1(M)$$ satisfy the left and right Leibniz rule, respectively, and they are known as {\it the induced quantum linear connections} (induced qlcs). The exterior covariant derivatives also fulfill $$d^{\nabla^{\omega}_{\alpha}}=\Upsilon_{\alpha}\circ D^{\omega}\circ \Upsilon^{-1}_{\alpha}\,,\quad d^{\widehat{\nabla}^{\omega}_{\alpha}}=\widehat{\Upsilon}_{\alpha}\circ \ast \circ D^{\omega}\circ \ast \circ \widehat{\Upsilon}^{-1}_{\alpha}.$$

In Differential Geometry, for a linear representation $\alpha$ of the Lie group of a principal bundle, the Gauge Principle (\cite{nodg}, \cite{sald}) gives us a characterization of vector--bundle--valued differential forms in terms of basic differential forms of type $\alpha$, which we can observe here with the fact that $\Upsilon_{\alpha}$, $\widehat{\Upsilon}_{\alpha}$ are isomorphisms. The Gauge Principle also tells us that induced linear connections and their exterior covariant derivatives on  associated vector bundles are given by covariant derivatives of principal connections. The theory presented shows the {\it non--commutative geometrical} counterpart of the Gauge Principle \cite{sald1}, \cite{sald2}.

Following the classical case, the inner products $\{\langle-|-\rangle_l\}$ and $\{\langle-|-\rangle_r\}$ can be extended to qvb--valued differential forms and
the maps $$d^{\nabla^{\omega}_{\alpha} \star_l }:=(-1)^{k+1} ((\star^{-1}_l\circ\ast)\otimes_M \id) \circ\; d^{\nabla^{\omega}_{\alpha}}\circ ((\ast\circ\star_l)\otimes_M \id), $$ $$d^{\widehat{\nabla}^{\omega}_{\alpha} \star_r}:=(-1)^{k+1} (\id\otimes_M (\star^{-1}_r\circ\ast)) \circ\; d^{\widehat{\nabla}^{\omega}_{\alpha}}\circ (\id\otimes_M (\ast\circ\star_r)). $$ are the adjoint operators of $d^{\nabla^{\omega}_{\alpha}}$ and $d^{\widehat{\nabla}^{\omega}_{\alpha}}$, respectively \cite{sald3}, i.e. $$\langle \psi'|d^{\nabla^{\omega}_{\alpha}}\psi\rangle_l=\langle d^{\nabla^{\omega}_{\alpha} \star_l }\psi'|\psi\rangle_l $$ $$\langle \varphi'|d^{\widehat{\nabla}^{\omega}_{\alpha}}\varphi\rangle_r=\langle d^{\widehat{\nabla}^{\omega}_{\alpha} \star_r}\varphi'|\varphi\rangle_r.$$

\section{The Quantum Gauge Group}

Let $\f_1,\,\f_2:\Gamma^\wedge\longrightarrow \Omega^\bullet(M\otimes G)$ be two graded--preserving linear maps. The convolution product of $\f_1$ with $\f_2$ is defined by $$\f_1\ast \f_2=m\circ (\f_1\otimes\f_2)\circ \phi:\Gamma^\wedge\longrightarrow \Omega^\bullet(M\otimes G),$$ where $m$ is the product of $\Omega^\bullet(M\otimes G)$. $\f$ is a convolution invertible map if there exists $\f^{-1}:\Gamma^\wedge\longrightarrow \Omega^\bullet(M\otimes G)$ such that $$\f\ast\f^{-1}=\f^{-1}\ast\f=\mathbbm{1}\epsilon;\qquad\f(\mathbbm{1})=\mathbbm{1};$$ 
and

\begin{equation*}
  \begin{CD}
  \Gamma^\wedge
   @>{\hbox to70pt{\hfil$\scriptstyle \Ad$\hfil}}>>
   &
   \Gamma^\wedge\otimes \Gamma^\wedge
   \\
   @V{\f}VV
   \hbox to0pt{\hskip-44pt$\circlearrowleft$\hss} & @VV{\f\,\otimes\,\id_{\Gamma^\wedge}}V
   \\[-4pt]
  \Omega^\bullet(M \otimes G)
   @>>{\hbox to70pt{\hfil$\scriptstyle \Psi $\hfil}}>
   &
   \Omega^\bullet(M \otimes G)\otimes \Gamma^\wedge,
  \end{CD}
 \end{equation*} 
where $\Ad:\Gamma^\wedge\longrightarrow\Gamma^\wedge\otimes\Gamma^\wedge$ is the extension of the right adjoint coaction $\Ad:G\longrightarrow G\otimes G$ \cite{woro2}. The quantum gauge group $GG$ is defined as the group of all convolution invertible maps \cite{sald2}.
$GG$ acts on the space of qpcs and on $\Gamma^l(M,V^\alpha M)$, $\Gamma^r(M,V^\alpha M)$ by means of $$F_\f\circ \omega\,,\qquad F_\f\circ T \,,\qquad \ast \circ F_\f\circ\ast\circ T,$$ respectively, where $$F_\f:=m\circ (\id_{\Omega^\bullet(M \otimes G)}\otimes \f)\circ \Psi.$$

\section{Yang--Mills Theory}

We can define the adjoint copresentation on ${_\inv}\Gamma$ as $$\ad(\varsigma)=\varsigma\otimes \mathbbm{1} ,$$ so we can consider the qvbs associated to $\ad$ . Since $R^\omega$, $\widehat{R}^\omega:=\ast\circ R^\omega\circ\ast$ $\in$ $\Mor(\ad, \Psi|_{\Hor^\bullet(M\otimes G)})$, the expression 
\begin{equation}
\label{f.4}
\L_{\YM}(\omega):=-\dfrac{1}{4}\left(\langle R^\omega,R^\omega\rangle_l +\langle \widehat{R}^\omega,\widehat{R}^\omega\rangle_r \right)
\end{equation}
is well--defined and it is called the {\it non--commutative geometrical Yang--Mills Lagrangian} \cite{sald3}. It is  gauge--invariant just for a subgroup $$GG_{\YM}:=\{\f\in GG| F_\f\circ \omega^\triv=\omega^\triv \mbox{ for all }\omega\} \supset S_2, $$ where $\omega^\triv(\varsigma)=\mathbbm{1}\otimes \varsigma$.

Critical points of its associated action
\begin{equation}
\label{f.5}
\mathscr{S}_{\YM}(\omega)=-\dfrac{1}{4}\left(\langle R^\omega|R^\omega\rangle_l +\langle \widehat{R}^\omega|\widehat{R}^\omega\rangle_r \right)
\end{equation}
satisfy
\begin{equation}
\label{7}
\langle \lambda | (d^{\nabla^{\omega}_{\ad} \star_l }-d^{S^{\omega} \star_l })R^\omega \rangle_l+\langle \ast \circ \lambda\circ \ast | (d^{\widehat{\nabla}^{\omega}_{\ad} \star_r }-d^{\widehat{S}^{\omega} \star_r })R^\omega \rangle_r=0,
\end{equation}
for all $\lambda$ $\in$ $\Mor^1(\ad,\Psi|_{\Hor^\bullet(M\otimes G)})$ (morphisms whose image lies on the space of {\it quantum} $1$--forms), where $d^{S^{\omega} \star_l }$, $d^{\widehat{S}^{\omega} \star_r }$ are the adjoint of the operators of $\Upsilon_\ad \circ S^{\omega}$, $\widehat{\Upsilon}_\ad \circ \ast \circ S^{\omega}\circ \ast,$ respectively. The operator $$S^{\omega}:\Mor(\ad,\Psi|_{\Hor^\bullet(M\otimes G)})\longrightarrow \Mor(\ad,\Psi|_{\Hor^\bullet(M\otimes G)})$$ in this case is given by $$S^{\omega}(\tau)(\varsigma)=2\,\left(\omega(\varsigma)\,\tau(\varsigma)-(-1)^k \tau(\varsigma)\,\omega(\varsigma)\right),$$ if $\tau$ $\in$ $\Mor^k(\ad,\Psi|_{\Hor^\bullet(M\otimes G)})$.

Solutions to the {\it non--commutative geometrical Yang--Mills equation} (eq.7) are flat qpcs in the sense that $R^\omega=0$, and $\omega_{\YM}$ with $$\omega_{\YM}(\varsigma)=\left(\begin{array}{lcr}
0  & \dfrac{i}{2} \\
\dfrac{i}{2} & 0 \end{array}\right)\otimes \mathbbm{1}+\mathbbm{1}\otimes \varsigma.$$The curvature of $\omega_{\YM}$ is given by $$R^{\omega_{\YM}}(\varsigma)=\left(\begin{array}{lcr}
-\dfrac{i}{2}  & 0 \\
\;\;\;0 & -\dfrac{i}{2} \end{array}\right) \;\in\; \Omega^2(M).$$
It is worth mentioning that the action of $GG_{\YM}$ on the set of critical points is trivial.

 In terms of a physical interpretation, {\it the Yang--Mills part of this example models a space--time of just two points with just one gauge boson field with symmetry $GG_{\YM} \supset S_2$, which is given by $\omega_{\YM}$.} However, it seems like the {\it vacuum} is full of non--gauge--equivalent boson fields (flat qpcs).

\section{Yang--Mills Scalar Matter Theory}

The non--commutative geometrical Yang--Mills--Scalar--Matter Lagrangian is given by \cite{sald3} $$\L_{\YMSM}(\omega,T_1,T_2):=-\dfrac{1}{4}\left(\langle R^\omega,R^\omega\rangle_l +\langle \widehat{R}^\omega,\widehat{R}^\omega\rangle_r \right)+$$ $$\dfrac{1}{4}\left( \langle \nabla^{\omega}_{\alpha}T_1,\nabla^{\omega}_{\alpha}T_1\rangle_l-V_l(T_1)-\langle \widehat{\nabla}^{\omega}_{\widehat{\alpha}}T_2,\widehat{\nabla}^{\omega}_{\widehat{\alpha}}T_2\rangle_r+V_r(T_2)\right),$$ where $\widehat{\alpha}$ is the conjugate corepresentation of $\alpha$ (\cite{woro1}, \cite{sald1}) and $V:M\longrightarrow M $ is a Fr\'echet differentiable map  called {\it the potential}; and $$V_l(T_1)= V(\langle T_1,T_1\rangle_l)\,,\quad V_2(T_2)= V(\langle T_2,T_2\rangle_r).$$

Critical points of its associated action
satisfy $$ \langle \Upsilon_{\alpha}\circ K^{\lambda}(T_1)\,|\,\nabla^{\omega}_{\alpha}T_1\rangle_l  -  \langle \widehat{\Upsilon}_{\widehat{\alpha}}\circ (\ast \circ K^{\lambda}\circ \ast) (T_2)\,|\,\widehat{\nabla}^{\omega}_{\widehat{\alpha}}T_2\rangle_r =$$
\begin{equation}
\label{7.f3.1}
\langle \lambda\,|\,(d^{\nabla^{\omega}_{\ad}\star_l}-d^{S^{\omega}\star_l}) R^{\omega}\rangle_l \; +\; \langle \ast \circ \lambda\circ \ast \,|\,(d^{\widehat{\nabla}^{\omega}_{\ad}\star_r}-d^{\widehat{S}^{q\omega}\star_r}) \widehat{R}^{q\omega}\rangle_r
\end{equation}
for all $\lambda$ $\in$ $\Mor^1(\ad,\Psi|_{\Hor^\bullet_{M\otimes G}})$, and
\begin{equation}
\label{7.f3.2}
\nabla^{q\omega\,\star_l}_{q\alpha}\left(\nabla^{\omega}_{\alpha}\,T_1\right)-V'_l(T_1)^\ast\,T_1=0,
\end{equation}
\begin{equation}
\label{7.f3.3}
 \widehat{\nabla}^{\omega\,\star_r}_{\widehat{\alpha}} \left(\widehat{\nabla}^{\omega}_{\widehat{\alpha}}\,T_2\right)-T_2 \,V'_r(T_2)^\ast=0,
\end{equation}
with $K^{\lambda}(T)=-T^{(0)}\lambda(\pi(T^{(0)}))$ for all $T$ $\in$ $\Mor(\alpha,\Phi)$ if $\Phi(T(v))=T^{(0)}(v)\otimes T^{(1)}(v)$ (in Sweedler's notation), where $\pi$ is the quantum germs map \cite{stheve}.

In Non--Commutative Geometry, it is necessary to take the left structure of $\Mor(\alpha,\Phi)$ and the right structure of  $\Mor(\widehat{\alpha},\Phi)$. In the {\it classical} case, the left and right structure are the same and $D^\omega=\ast \circ D^\omega \circ \ast$, so our theory recreates the classical case  (Equations (3) and (4) for a {\it real} potential) but considering antiparticles too. Therefore,  Equations (8), (9) and (10) are the {\it non--commutative geometrical version or a non--commutative geometrical generalization} of Equations (3), (4).

To calculate solutions for these equations, we have to choose the correpresentation and the potential $V$. For example, for the trivial corepresentation on $\C$ and a potential $V$ such that for all $p$ $\in$ $M$  $$V'(p)=\left(\begin{array}{lcr}
 2-2\dfrac{y}{x}             &  0 \\
 0 & 2-2\dfrac{x}{y} \\
\end{array}\right)$$ for some fixed $x$, $y$ $\in$ $\R$, critical points are triplets $(\omega,T_1,T_2)$, where $\omega$ is a flat qpc or $\omega_{\YM}$ and $T_1$ $\in$ $\Gamma^l(M,\C M)$, $T_2$ $\in$ $\Gamma^r(M,\C M)$ are defined by $$T_1(1)=T_2(1)=\left(\begin{array}{lcr}
 x             &  0 \\
 0 & y \\
\end{array}\right).$$

In this case, we can ensure that at least, the Lagrangian is gauge--invariant for $GG_\YM$. 

 In terms of a physical interpretation, {\it this example models left scalar matter and right scalar antimatter fields in the trivial corepresentation  coupled to gauge boson fields in a space--time of just two points.}

Now let us consider the alternanting quantum representation on $\C$, $\alpha^\altt$. In this case a left--right $M$ basis of $\Mor(\alpha^\altt,\Phi) $ is given by the element
\begin{equation*}
\begin{aligned}
T^\altt:\C &\longrightarrow M\otimes G \\
\lambda &\longmapsto  \lambda\mathbbm{1} \otimes \mathbbm{1}^\altt 
\end{aligned}
\end{equation*}
where $\mathbbm{1}^\altt=\Delta_0-\Delta_1$ and hence, every $T$ $\in$ $\Mor(\alpha^\altt_\C,\Phi)$ is of the form $T=p^{T}T^\altt=T^\altt\, p^{T} $ where $p^{T}=T(1)(\mathbbm{1} \otimes \mathbbm{1}^\altt)$.\\ 

In general, for a qpc $\omega$ with $q\omega(\varsigma)=\left(\begin{array}{lcr}
\,0    &  \lambda_0 \\
\lambda_1 & 0 \\
\end{array}\right)\otimes \mathbbm{1}+\mathbbm{1}\otimes \varsigma$ and $T_1=\left(\begin{array}{lcr}
\widetilde{p}_0    &  0 \\
0 & \widetilde{p}_1 \\
\end{array}\right)T^\altt$, $T_2=T^\altt \left(\begin{array}{lcr}
\hat{p}_0    &  0 \\
0 & \hat{p}_1 \\
\end{array}\right)$, Equation \ref{7.f3.1} turns into 
\begin{equation}
\label{7.f3.4}
\begin{split}
i(|\widetilde{p}_0|^2-\widetilde{p}_0\widetilde{p}^\ast_1+\hat{p}^\ast_0\hat{p}_1-|\hat{p}_0|^2)+2(|\widetilde{p}_0|^2-|\hat{p}_0|^2))\lambda^\ast_0=u^\ast (1+2i\lambda_1)\\
i(|\widetilde{p}_1|^2-\widetilde{p}^\ast_0\widetilde{p}_1+\hat{p}_0\hat{p}^\ast_1-|\hat{p}_1|^2)+2(|\widetilde{p}_1|^2-|\hat{p}_1|^2))\lambda^\ast_1=u^\ast (1+2i\lambda_0),
\end{split}
\end{equation}
where $u=-(\lambda_0+\lambda_1)-2\,i\,\lambda_0\,\lambda_1$; while Equation \ref{7.f3.2} turns into 
\begin{equation}
\label{7.f3.5}
\begin{aligned}
\nabla^{\omega\,\star_l}_{\alpha^\altt_\C}\left(\nabla^{\omega}_{\alpha^\altt_\C}\,T_1\right)=\left(\begin{array}{lcr}
\widetilde{u}_0 & 0  \\
0 & \widetilde{u}_1
\end{array}\right)T^\altt,\\ \widehat{\nabla}^{\omega\,\star_r}_{\alpha^\altt_\C} \left(\widehat{\nabla}^{\omega}_{\alpha^\altt_\C}\,T_2\right)=T^\altt\left(\begin{array}{lcr}
\widehat{u}_0 & 0  \\
0 & \widehat{u}_1
\end{array}\right),
\end{aligned}
\end{equation}
where $\widetilde{u}_0:=2(\widetilde{p}_0-\widetilde{p}_1)+2i\widetilde{p}_0(\lambda_0+\lambda_1)-4i\widetilde{p}_1\lambda_1-4\widetilde{p}_0\lambda_0\lambda_1$,  $\widetilde{u}_1:=-2(\widetilde{p}_0-\widetilde{p}_1)+2i\widetilde{p}_1(\lambda_0+\lambda_1)-4i\widetilde{p}_0\lambda_0-4\widetilde{p}_1\lambda_0\lambda_1$, $\hat{u}_0:=2(\hat{p}_0-\hat{p}_1)+2i\hat{p}_0(\lambda_0+\lambda_1)-4i\hat{p}_1\lambda_0-4\hat{p}_0\lambda_0\lambda_1 $, $\hat{u}_1:=-2(\hat{p}_0-\hat{p}_1)+2i\hat{p}_1(\lambda_0+\lambda_1)-4i\hat{p}_0\lambda_1-4\hat{p}_1\lambda_0\lambda_1.$ This allows us to find solutions, for example, taking $\omega_\YM$ and any $T_1$, $T_2$, Equation \ref{7.f3.3} turns into $$\widetilde{p}_0\widetilde{p}^\ast_1= \hat{p}^\ast_0\hat{p}_1;$$ while Equation \ref{7.f3.4} turns into $$ \nabla^{q\omega\,\star_l}_{q\alpha^\altt_\C}\left(\nabla^{q\omega}_{q\alpha^\altt_\C}\,T_1\right)=T_1\,,\qquad \widehat{\nabla}^{q\omega\,\star_r}_{q\alpha^\altt_\C} \left(\widehat{\nabla}^{q\omega}_{q\alpha^\altt_\C}\,T_2\right)=T_2.$$ Of course, there are more solutions. 

Finally it is easy to show that 
\begin{equation*}
(d^{\nabla^{q\omega}_{\ad}\star_l}-d^{S^{q\omega}\star_l})^2=(d^{\widehat{\nabla}^{q\omega}_{\ad}\star_r}-d^{\widehat{S}^{q\omega}\star_r})^2=0;
\end{equation*}
so in this case a kind of {\it left--right noncommutative continuity equation} holds.\\

At least we can ensure that $$\{\f\in GG \mid \f(\mathbbm{1}^\altt)=e^{it}\mathbbm{1}\;, F_\f \circ \omega^\triv =\omega^\triv \; \mbox{ with }t\,\in\, \R\} $$  is a subgroup of $GG$ for which the Lagrangian is always gauge--invariant, no matter the form of the potential $V$. 

In terms of a physical interpretation, {\it  this example models left scalar matter and right scalar antimatter fields in the alternanting representation coupled to gauge boson fields in a space--time of just two points.}

\section{Final Comments}

To conclude this work we have to talk about some interesting points

\begin{enumerate}
\item There is a general theory that supports all the concepts presented. This was just an example.
\item The operator $S^{\omega}$ does not have a {\it classical} counterpart in the sense that in Differential Geometry, $S^{\omega}=0$. Its participation in Equations (7) and (8) is completely a {\it quantum phenomena}.
\item Our definition of the quantum gauge group is based on the one presented by T. Brzezinski in \cite{Br}; nevertheless, it does not recreate the {\it classical} case. One option is considering the group generated by all convolution invertible maps $\f$ that also are graded differential $\ast$--algebras morphisms. Another option is considering the group of all maps $\F_\f$ such that they are graded differential $\ast$--algebra difeomorphisms. However, applying these ideas in other examples, there are not enough gauge transformations.
\item Although all these constructions are developed in the framework of Non--Commutative Geometry, the quantum gauge group is a {\it classical} group. Therefore, an exciting way of research would be trying to define the quantum gauge group as a quantum group.
\item In the Lagrangian $\L_{\YMSM}$ we work with left and right structures and  with a corepresentation and its conjugate. In the {\it quantum case} both parts are necessary and this can be appreciated when we work with the {\it quantum Hopf fibration} \cite{sald4}. In this case, considering both parts one can easily find solutions. It looks like if in the  {\it quantum case left particles and right antiparticles cannot be separated}; they appear naturally interconnected. 
\end{enumerate}


\begin{thebibliography}{99999}
%
 \bibitem[Br]{Br}
  \textsc{Brzezinski, T~:}\quad
  \textit{Translation Map in Quantum Principal Bundles,\ }
  \textrm{J. Geom. Phys. 20, 349 (1996).}
%
 \bibitem[D1]{micho1}
  \textsc{Durdevich, M.~:}\quad
  \textit{Geometry of Quantum Principal Bundles I,\ }
  \textrm{Commun.~Math.~Phys.~{\bf 175} (3), 457---521 (1996).}  
%
 \bibitem[D2]{micho2}
  \textsc{Durdevich, M.~:}\quad
  \textit{Geometry of Quantum Principal Bundles II,\ }
  \textrm{Rev.~Math.~Phys.~{\bf 9} (5), 531---607 (1997).}
 %
\bibitem[KMS]{nodg} \textsc{Kol\'ar, I., Michor, P. W. \& Slovák, J.~:}
\quad
  \textit{Natural Operations in Differential Geometry,\ } internet book 

http://www.mat.univie.ac.at/~michor/kmsbookh.pdf   
%
 \bibitem[Sa11]{sald1}
  \textsc{Salda\~na, M, G.~:}\quad
  \textit{Functoriality of Quantum Principal Bundles and Quantum Connections,\ }
  \textrm{arXiv:2002.04015, 10 Feb 2020.} 
%
 \bibitem[Sa12]{sald2}
  \textsc{Salda\~na, M, G.~:}\quad
  \textit{Geometry of Associated Quantum Vector Bundles and The Quantum Gauge Group.\ }
  \textrm{arXiv:2109.01550v1, 3 Sep 2021}
%
 \bibitem[Sa13]{sald3}
  \textsc{Salda\~na, M, G.~:}\quad
  \textit{Qauntum Principal Bundles and Yang--Mills--Scalar--Matter Fields.\ }
  \textrm{arXiv:2109.01554v1, 3 Sep 2021}
  %
 \bibitem[Sa14]{sald4}
  \textsc{Salda\~na, M, G.~:}\quad
  \textit{Yang--Mills--Scalar--Matter Fields in the Quantum Hopf Fibration.\ }
  \textrm{In preparation.}
%
  \bibitem[SW]{sald}
  \textsc{Salda\~na, M, G, A. \& Weingart, G.~:}\quad
  \textit{Functoriality of Principal Bundles and Connections,\ }
  \textrm{arXiv:1907.10231v2, 18 Apr 2020.}
%
 \bibitem[So]{stheve}
  \textsc{Sontz, S, B.~:}\quad
  \textit{Principal Bundles: The Quantum Case,\ }
  \textrm{Universitext, Springer, 2015.}
%
  \bibitem[W1]{woro1}
  \textsc{Woronowicz, S, L.~:}\quad
  \textit{Compact Matrix Pseudogroups,\ }
  \textrm{Commun. Math. Phys. {\bf 111}, 613-665 (1987).}
  %
 \bibitem[W2]{woro2}
  \textsc{Woronowicz, S, L.~:}\quad
  \textit{Differential Calculus on Compact Matrix Pseudogroups (Quantum Groups),\ }
  \textrm{Commun. Math. Phys. {\bf 122}, 125-170 (1989).}
\end{thebibliography}
\end{document}